\documentclass[12pt,preprint]{aastex}
\newcommand{\Msolar}{\mbox{\,$\rm M_{\odot}$}}        % solar mass
\begin{document}

\title
{TIMESCALES OF SOFT X-RAY VARIABILITY AND PHYSICAL CONSTRAINTS IN
AGNs}

\author
{W.Bian\altaffilmark{1,2} \and Y.Zhao\altaffilmark{1}}

\altaffiltext{1}{National Astronomical Observatories, Chinese
Academy of Sciences, Beijing 100012, China}
\altaffiltext{2}{Department of Physics, Nanjing Normal University,
Nanjing 210097, China}

%\date{Received ...; accepted ...}

\begin{abstract}
We present soft X-ray variability time-scales for 65 active
galactic nuclei (AGNs) derived from ROSAT/PSPC pointing data. For
these 65 objects with obvious exponential increasing or decreasing
patterns in their light curves, we use the exponential time-scales
and find they are more suitable for analyzing ROSAT light curve
data. We also discuss some physical constraints on the central
engine implied by our results. Assuming that this soft X-ray
variability exponential time-scale is approximately equal to the
thermal timescale of the standard accretion disk, we obtain the
accretion rate, the size of soft x-ray radiation region and the
compactness parameter for 37 AGNs, using their recently estimated
central black hole masses. For 12 of these 37 AGNs, the radii of
the gravitational instability in the standard thin accretion disks
are obtained using central black hole masses and the calculated
accretion rates. These are consistent with the results from the
reverberation mapping method. These results provide supporting
evidence that such gravitational instabilities contribute to the
formation of the Broad Line Regions (BLRs) in AGNs.

\end{abstract}


\section{INTRODUCTION}

X-ray variability has long been known to be a common property of
active galactic nuclei (AGNs). The X-ray variability in the
continuum and in the emission lines can give a clue to the central
engine in AGNs (Mushotzky et al. 1993). The variance amplitude and
the time-scale are two main parameters which characterize the
variance. There are many quantities can be used to describe the
variability amplitude or the timescale in X-ray variability, such
as the flux-doubling time-scale (Barr \& Mushotzky 1986),
normalized variability amplitude (Green et al. 1993), and the
excess variance (Nandra et al. 1997). Some powerful methods are
also introduced in the analysis of the X-ray light curves. These
include the power density spectrum (PDS), the wavelet method
(Otazu et al. 2002 ), and the power density in the time domain (Li
2001).

There has been steady progress on the analysis of X-ray
variability since the launch of EXOSAT, RXTE, ROSAT, ASCA,
XMM-Newton. For an individual X-ray light curve, PDS analysis is
now popular and can offer more information about the X-ray
variability. Early attempts to constrain a PDS flattening or ``cut
off'', which relates to a characteristic time-scale, were made
with the EXOSAT data of NGC 5506 (McHardy 1988) and NGC 4151
(Papadakis 1995). These analyses yield evidence for a cutoff
time-scale of several weeks. However, the uneven sampling of these
data made their reliability uncertain. The situation was
significantly improved with the launch of RXTE. Using RXTE data,
Edelson \& Nandra (1999) obtained an evenly sampled X-ray light
curve of NGC 3516 and produced a PDS covering four decades in
temporal frequency, finding a cutoff time-scale of about one
month. Pounds et al. (2001) also found a cutoff time-scale of
about 13 days for a narrow line Seyfert 1 galaxy (Akn 564). Uttley
et al. (2002) recently developed a Monte Carlo method to test
models for the true power spectrum of intermittently sampled light
curves against the observed noisy power spectra of four Seyfert
galaxies, which are flat towards lower frequencies. They fit two
models for the flattening: a ``knee'' model and a ``high-frequency
break'' model. They reduced the characteristic frequency for these
four AGNs.

The X-ray characteristic time-scale or frequency usually relates
to the size of the X-ray emission region, which then can be used
to constrain the central engine, e.g. the central black hole
masses, accretion rates, and the radiation mechanism.

ROSAT observations of AGNs offer plenty of information about the
soft X-ray variability, which is thought to come from the
accretion disk around the central black hole in these AGNs.
However, PDS analysis is not suitable for analyzing ROSAT pointed
data, since there are usually large time gaps between adjacent
exposures in ROSAT pointing observations. Thus, time-scale
analysis remains necessary in AGN samples. There are obvious
exponentially increasing or decreasing patterns in the soft X-ray
light curves in some AGNs observed with ROSAT/PSPC in pointing
mode. In this paper we use the exponential time-scale (Zhao \&
Fink 1996) to investigate the soft X-ray variability in AGNs with
this kind of pattern. This paper is a continuation of the previous
work of Zhao \& Fink (1996) .

The standard model of AGNs is an accretion disk with a jet around
a central massive black hole. Progress on a reliable estimate
about black hole masses can effectively constrain the central
engine when combined with the research on the X-ray variability.
Several methods have been used to estimate the mass of a black
hole in the center of a galaxy. These include stellar dynamical
studies (reviewed by Kormendy \& Gebhardt 2001) , reverberation
mapping methods (Kaspi et al. 2000), analyses of the high
frequency tail of the power density spectrum (Czerny et al. 2001),
the relation between central black hole masses and the bulge
velocity dispersion $M_{BH}-\sigma$ (Merritt \& Ferrarese 2001),
and the use of single-epoch rest-frame optical spectrophotometric
measurements (Vestergaard 2002).

In next section time-scales and central black hole masses of our
sample are presented. We then give the physical constraints on the
central engine of 38 AGNs in Section 3. In Section 4 we derive the
radii of the gravitational instability in the standard $\alpha$
accretion disks of 12 AGNs and compare them with BLRs sizes from
the reverberation mapping method. Our results are summarized in
Section 5. The cosmological parameters $H_{0}=\rm 75 km s^{-1}
Mpc^{-1}$and $q_{0}=0.5$ have been adopted in this work.

\section{SAMPLE AND TIMESCALES OF VARIABILITY}

All sources were observed in ROSAT/PSPC mode over periods varying
from days to years. We picked out the AGNs from cross
identification of the Veron-Cetty (2001) AGN catalogue with the
ROSAT point source catalog. From the ROSAT public archive of PSPC
observations, only sources with total X-ray photon counts greater
than 1000 were selected to ensure the quality of the X-ray
spectra. This yielded more than 200 AGNs. The data were processed
for instrument corrections (such as vignetting and dead time
effects) and background subtraction using the EXSAS/MIDAS
software.

The light curve for each AGN was obtained from original ROSAT
observations with a 400 seconds time bin in three energy bands;
0.1-2.4 (total band), 0.1-0.4 (A band), 0.5-2.0 (B band) keV. The
estimates of time-scales for all AGNs described below are based on
those light curves. From these light curves we can find that there
exist obvious exponentially increasing or decreasing patterns.
Since X-ray emission is believed to come from the accretion disk
around the central black hole in AGNs, these exponential patterns
may provide physical insight. We select the AGN light curves with
exponential patterns by the following criteria: (1)The count rate
increased or decayed continually in the light curves, considering
count rate uncertainties of 3 $\sigma$, (2)the number of the
contiguous data points fitted by the exponential function (Eq.
\ref{te}) are greater than 4, (3)the magnitude of the count rate
variance is greater than $5 \sqrt {(\sigma_{1}^2 + \sigma
_{2}^2)}$, where $\sigma_{1}$ and $\sigma_{2}$ are the count rate
errors of the jumping-off point and end-point, respectively. The
soft X-ray luminosity in the 0.1-2.4 keV of each AGN is obtained
by averaging the power-law fitting spectra of all the
observations.

\subsection{Exponential Time-scales}
We use the exponential time-scale (Zhao \& Fink 1996) to
characterize the exponential pattern in the light curves in AGNs.
The exponential time-scale, $\Delta t_{e}$, is defined as
\begin{equation}
\label{te} I(t)=I_{0}+I_{a}e^{\pm \frac{t}{\Delta t_{e}}}.
\end{equation}
where $I(t)$ is the count rate at t, $I_{0}$ and $I_{a}$ are
constants.

\begin{figure}
%\centerline{\epsfig{file=tm0.ps,width=10cm,angle=90,clip=}}
\plotone{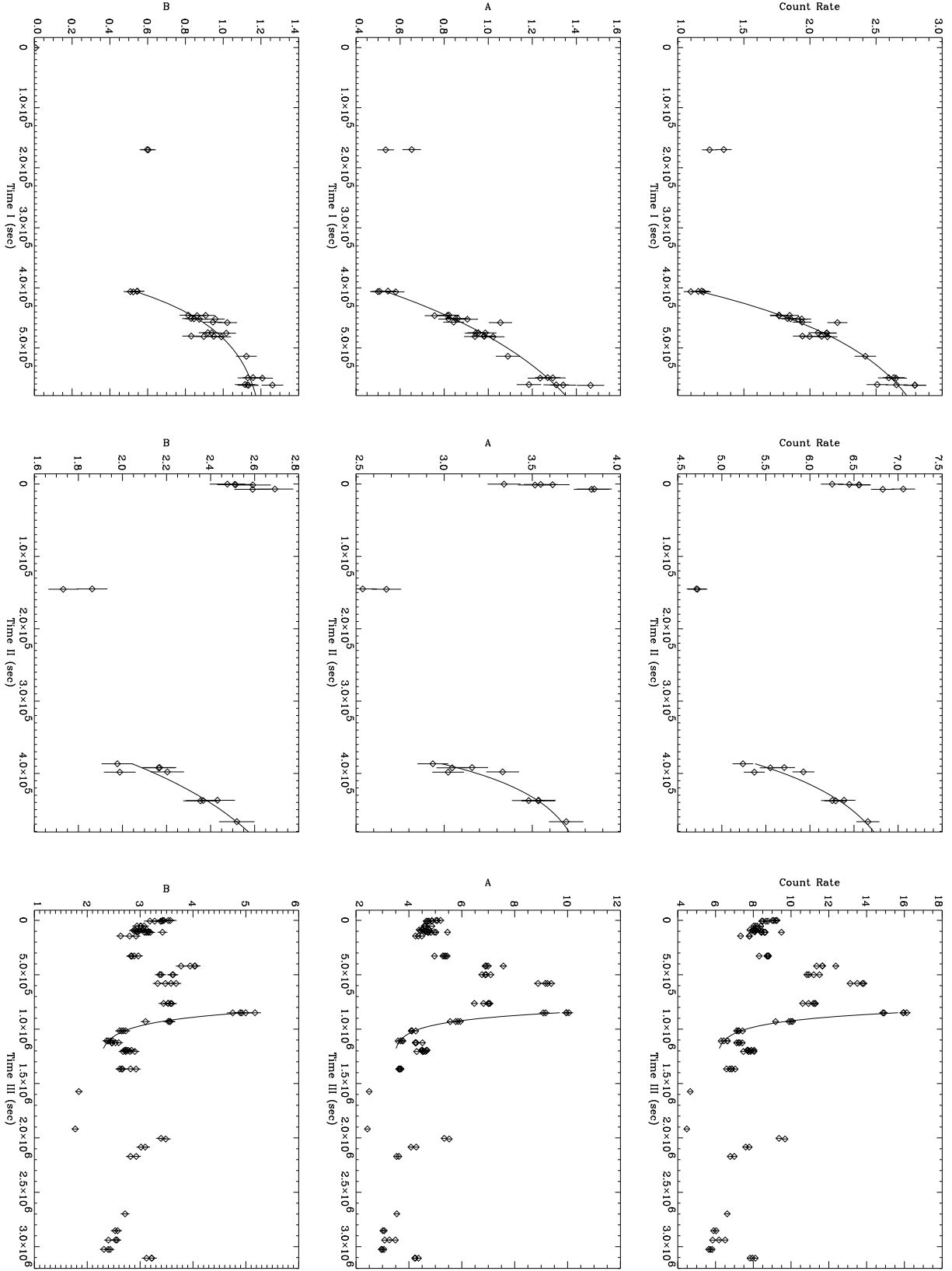}

\caption{Variability of count rate and exponential time-scale
fitting of NGC 5548 in three energy band: 0.1-2.4 (total band),
0.1-0.4 (A band), 0.5-2.0 (B band) keV. The time zero is at
22:48:59 UT in 1990 July 18 for Time I, at 23:19:29 UT in 1992
January 17 for Time II, and at 9:23:14 UT in 1992 December 24 for
Time III. }

\label{cr-te}
\end{figure}

The definition of exponential time-scale is based on instabilities
which increase or decrease exponentially. Thus the exponential
time-scale is equivalent to a rising or a descending time-scale
and $I_{a}$ is the amplitude of variability. $I_{0}$ is an
additional constant component, which is equal to the intensity
before or after the instability. It is exponentially increasing if
the sign of $I_{a}$ is positive while it is exponentially decaying
if the sign is negative. For AGNs with some values of $\Delta
t_{e}$, we select $\Delta t_{e}$ with the smallest error as the
time-scale. Taking the Seyfert 1 galaxy NGC 5548 as an example
(Fig. \ref{cr-te}), there are three obvious exponentially
increasing or decreasing patterns in the light curve. Three
patterns fitting by Eq. \ref{te} showed similar exponential
time-scales. The exponential time-scales are
$(1.04\pm0.32)\times10^{5}$, $(6.65\pm5.35)\times10^{4}$,
$(8.17\pm0.26)\times10^{4}$ seconds, respectively, which are
consistent if taking the errors into account. And $I_{0}$ are
$3.10\pm 0.13$, $7.16\pm 0.87$, $6.01\pm0.10$ photons $s^{-1}$,
respectively.

We think there is more physical information in the exponential
time-scale than in the two-folding time-scale, which only
considers the amplitude increased by a factor of two. There are
usually large time gaps between adjacent exposures in ROSAT
pointed observations. The two-folding time-scale will overestimate
the time-scale of ROSAT sources. Using our selection criteria as
defined above, we searched for any obvious continually
exponentially increasing or decreasing pattern in the light curve
of each AGN and then fitted the pattern with the form of Eq.
\ref{te}, considering the errors of the count rates (Press et al.
1992). The exponential time-scales fir 65 AGNs are listed in
column 3 of Table 1, in which there are nine QSOs, 37 Seyfert 1
galaxies, nine NLS1s, one Seyfert 2 galaxies, four BL Lacs and
five high optical polarization (HP) objects. The time-scales
calculated as rising exponentials are marked with a star. The rest
are derived from decaying exponentials.

\subsection{Estimations of Central Black Hole Masses}
\label{mass}

The masses of 13 AGNs labelled with ``a'' in Table 1 are adopted
from Nelson (2000). They were estimated from the reverberation
mapping methods with a high level of reliability. There is strong
correlation between the black hole mass and the host velocity
dispersion, namely the $M_{BH}-\sigma$ relation (Ferrarese et al.
2001). We adopt the $M_{BH}-\sigma$ relation found by Merritt \&
Ferrarese (2001), namely,

\begin{equation}
\label{Mcal} M_{BH}=1.3 \times 10^{8} \Msolar (\sigma/\rm {200 km
s^{-1}})^{4.72}.
\end{equation}

We derive masses for seven AGNs with measured host velocity
dispersions, which are labelled with ``b'' in Table 1 (Nelson \&
Whittle 1995; Falomo et al. 2002). The errors for black hole
masses in these seven AGNs are estimated from the error of the
bulge velocity dispersion. An empirical relationship between
single-epoch rest-frame optical spectrophotometric measurements
and the central masses was recently presented (Vestergaard 2002).
These methods for estimating AGN masses are all calibrated using
masses from the reverberation mapping method. The total number of
AGNs with available black hole masses is 37, of which there are 22
Seyfert 1 galaxies, five NLS1s, one Seyfert 2 galaxies, eight
Quasars and one BL Lac objects. The central black hole masses of
these 37 AGNs are listed in column 5 in Table 1.

\section{ANALYSIS AND PHYSICAL CONSTRAINTS}

\subsection{Central Black Hole Mass and Mass Limits from the exponential Time-scale}

The upper limit on the mass of the central black hole in AGNs
derived from the time-scale is defined as (Mushotzky et al. 1993):
\begin{equation}
\label{Mte} M(\Delta t)=2\times 10^{4}(\frac{\Delta
t}{sec})\Msolar.
\end{equation}
assuming that the size of the X-ray radiation region is 5
Schwarzschild radii.

\begin{figure}
\plotone{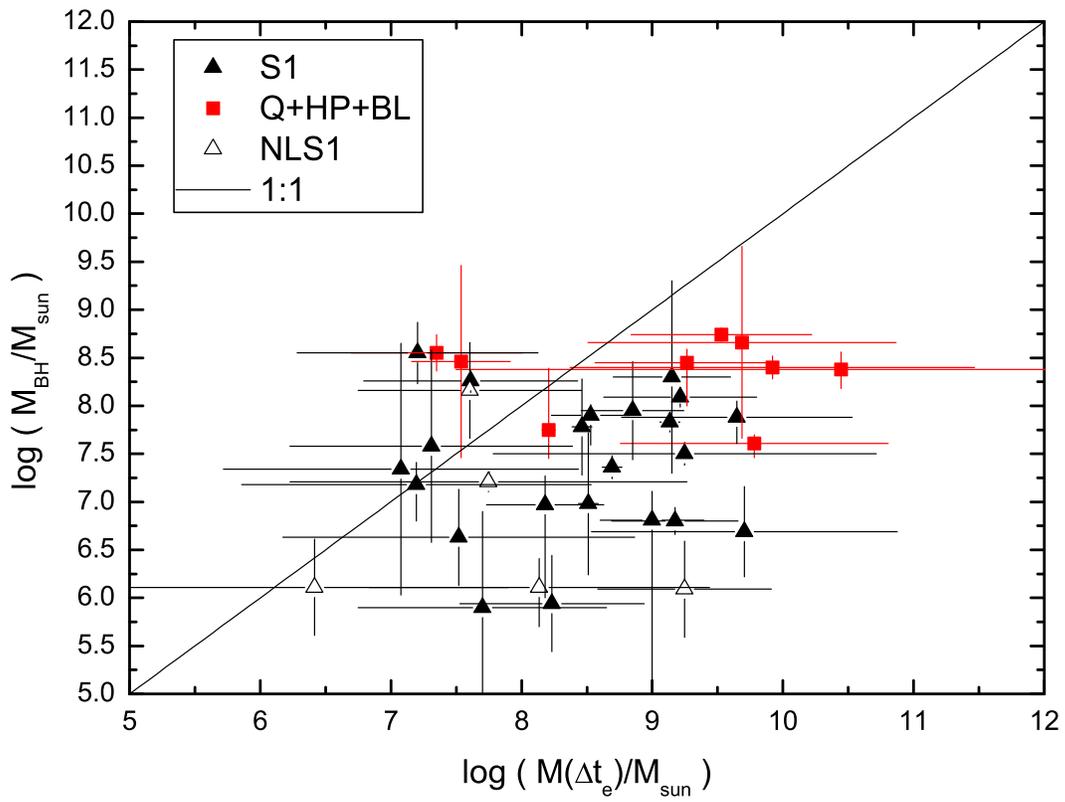}
%\centerline{\epsfig{file=M-Mt.eps,width=10cm,angle=0,clip=}}
\caption{The black hole masses versus the mass limits derived from
the exponential time-scales. The solid line shows 1:1.}
\label{M-Mt}
\end{figure}

We compared this mass limit derived from the exponential
time-scales with the above estimated central black hole masses
(see Sec. \ref{mass}) in Figure \ref{M-Mt}. This shows that the
exponential time-scale can also provide an effective limit for the
mass of a central black hole in an AGN.

\subsection{Constrains on the Radiation Mechanism in AGNs}

Some physical constraints on the relationship of bolometric
luminosity and the time-scale were proposed by Fabian (1979;
1992). If the source luminosity is below the Eddington limit, we
obtained a limit:
\begin{equation}
L_{bol}\leq \frac{2}{3}\pi \frac{m_{p}c^{4}}{\sigma_{T}}\Delta t
~~~~~\rm {ergs ~~s^{-1}},
\end{equation}
When the luminosity is produced by matter being transferred into
radiation with some efficiency $\eta\ll 1$, there is a second
limit:
\begin{equation}
L_{bol}\leq \eta \frac{m_{p}c^{4}}{\sigma_{T}}\Delta t ~~~~~\rm
{ergs ~~s^{-1}},
\end{equation}
If the primary spectrum of the source extends to $\gamma$-ray
energies, electron-positron pair production gives a third limit:
\begin{equation}
L_{bol}\leq 20 \pi \frac{m_{e}c^{4}}{\sigma_{T}}\Delta t ~~~~~\rm
{ergs ~~s^{-1}},
\end{equation}
Thermal bremsstrahlung radiation supplies more power than Compton
process only if:
\begin{equation}
L_{bol}\leq 10^{39} T_{9}^{-3/2}\Delta t ~~~~~\rm {ergs ~~s^{-1}}.
\end{equation}
where the temperature of the electrons is $T_{9}$ $10^{9}$K.

\begin{figure}
\plotone{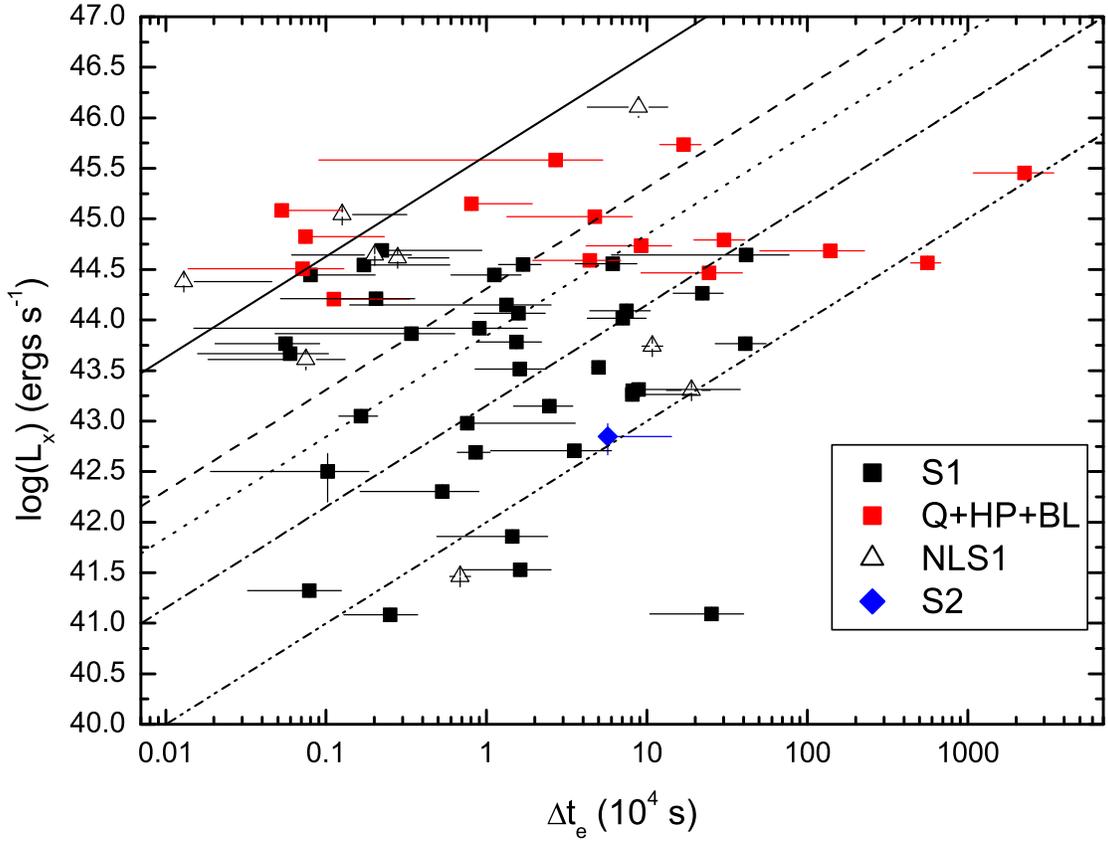}
%\centerline{\epsfig{file=lx-te.eps,width=10cm,angle=0,clip=}}
\caption{The soft X-ray luminosity versus the exponential
time-scales. Solid line: Eddington limit; dash line: efficiency
limit ($\eta$=0.1) ; dot line: compact parameter limit (l=60);
dash dot line: efficiency limit ($\eta$=0.007) ; dash dot dot
line: bremsstrahlung limit, respectively, assuming
$L_{bol}=10L_{x}$. } \label{lx-te}
\end{figure}

The bolometric luminosity is larger than the X-ray luminosity in
ROSAT's energy band. For example, the ratio of the bolometric
luminosity of 3C 273 (Padovani \& Rafanelli 1988), $1.7\times
10^{47} \rm{ergs ~~s^{-1}}$, and the X-ray luminosity in ROSAT's
energy band, $1.2\times 10^{46} \rm{ergs ~~s^{-1}}$, is 14. In
Figure \ref{lx-te} we show the soft X-ray luminosity in 0.1-2.4
keV versus the exponential time-scale. The lines in Figure
\ref{lx-te} are from Eq. 4-7 assuming $L_{bol}=10L_{x}$.

From Figure \ref{lx-te} bremsstrahlung is ruled out as the primary
radiation source except in several low luminosity sources because
of the observed rapid variability.

The objects whose bolometric luminosity are larger than the
Eddington limit are MS 01585+0019 (BL) ,PKS 0548-322 (HP) ,MS
23409-1511 (NLS1), PG 1244+026 (NLS1) . These objects may display
relativistic effects.

The efficiency factors, $\eta$, of most objects are larger than
0.007, which would not be associated with stellar process (see
Fig.~\ref{lx-te}) . It suggests strongly that most AGNs, at least,
have non-thermal and non-stellar processes which are produced by a
very efficient central engine.

\begin{figure}
\plotone{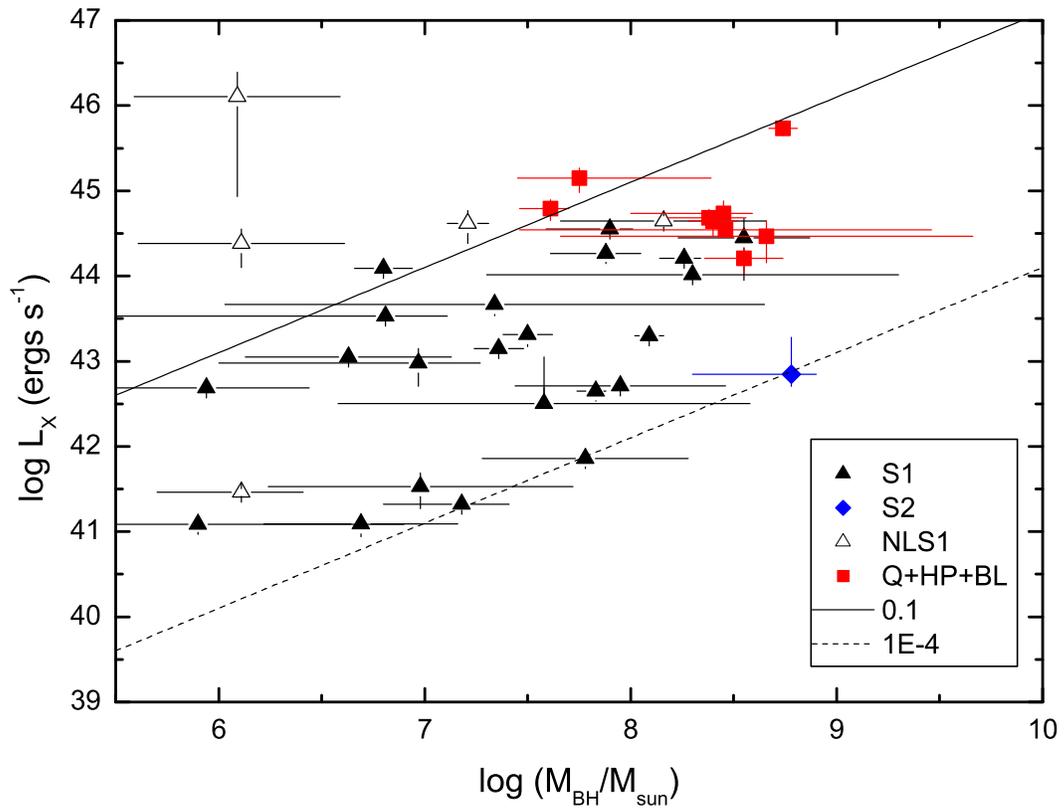}
%\centerline{\epsfig{file=lx-m.eps,width=10cm,angle=0,clip=}}
\caption{The soft X-ray luminosity versus the black hole masses.
The solid line shows $L_{x}=0.1L_{Edd}$. and dash line is
$L_{x}=10^{-4}L_{Edd}$.} \label{lx-m}
\end{figure}

We also plot $L_{x}$ versus central black hole masses in Figure
\ref{lx-m}. The solid line shows $L_{x}=0.1L_{Edd}$. and the dash
line is $L_{x}=10^{-4}L_{Edd}$. If we assume $L_{bol}=10L_{x}$, we
find that the efficiency factor is between 1 to $10^{-3}$, which
is consistent with the results from $L_{x}-\Delta t$ diagram ( see
Fig.~\ref{lx-te}).

\subsection{Sizes of the X-ray Radiation Regions }

Under unified model schemes, an accretion disk surrounds the
central massive black hole in an AGN. There are two instabilities,
thermal and viscous, usually occurring in accretion disks.
However, the rapid variability in X-rays cannot arise from viscous
instability because the viscous time-scale is too long (Mushotzky
et al. 1993). The time-scale of thermal instability in accretion
disks (Frank et al. 1992) is

\begin{equation}
t_{th}\sim \frac{1}{\alpha \Omega_{k}},
\end{equation}

where $\alpha$ is a parameter in the standard thin accretion disk,
$\Omega_{k}=\sqrt{\frac{GM}{R^{3}}}$ is the Keplerian angular
velocity in the orbit of radius R surrounding a central black hole
with the mass of M.

Assuming the exponential time-scale is approximately equal to the
thermal time-scale, we can obtained the size of radiation region,
$R_{x}$, as
\begin{equation}
\frac{R_{x}}{R_{g}}=(\frac{\alpha}{\sqrt{2}}\frac{c\Delta t_{e}
}{R_{g}})^{2/3}.
\end{equation}

where $R_{g}=\frac{2GM}{c^{2}}$ is the Schwarzschild radius.
$\alpha=1$ is adopted.

\begin{figure}
\plotone{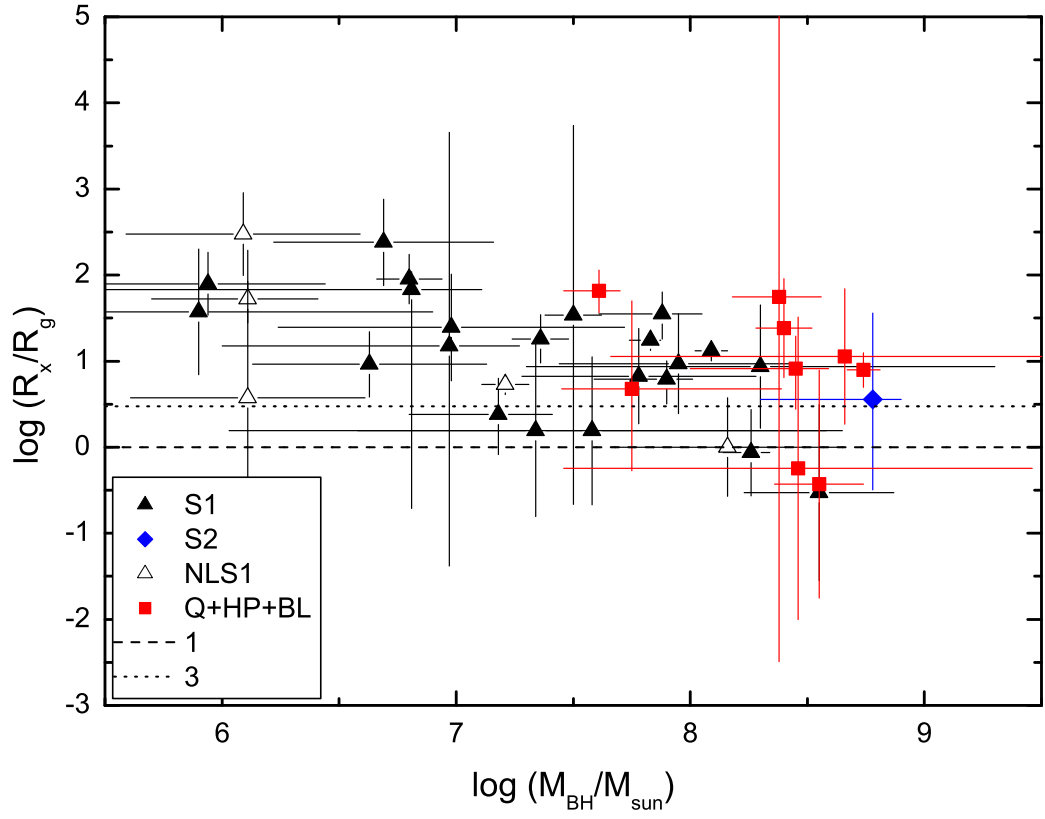}
%\centerline{\epsfig{file=rx-m.eps,width=10cm,angle=0,clip=}}
\caption{$log(R_{x}/R_{g})$ versus $log(M_{BH})$. The dash line is
$R_{x}/R_{g}$=1 and the dot line is $R_{x}/R_{g}$=3.} \label{rx-m}
\end{figure}

Taking the above estimated black hole masses and exponential
time-scales, we can obtained the ratios of the size of X-ray
radiation region to the Schwarzschild radius, which are listed in
column 7 of Table 1. $R_{x}/R_{g}$ distribution is
$<log(R_{x}/R_{g})>=0.97\pm 0.12$ with a standard deviation of
0.75. We plot $R_{x}/R_{g}$ versus $M_{BH}$ in Figure \ref{rx-m}.
A simple least square linear regression (Press et al. 1992) gives
$log(R_{x}/R_{g})=4.52\pm 0.92+(-0.47\pm0.12)log(M_{BH}/\Msolar)$,
with a Pearson correlation coefficient of R=0.54 corresponding to
a probability of P$<$0.000431 that the correlation is caused by a
random factor. The error of $R_{x}/R_{g}$ is estimated from the
errors of the exponential time-scale and the black hole mass. The
sizes of radiation regions of all AGNs are larger than the
so-called last stable radius $3R_{g}$, if we consider errors of
calculated sizes.

\subsection{Accretion Rates}

\begin{figure}
\plotone{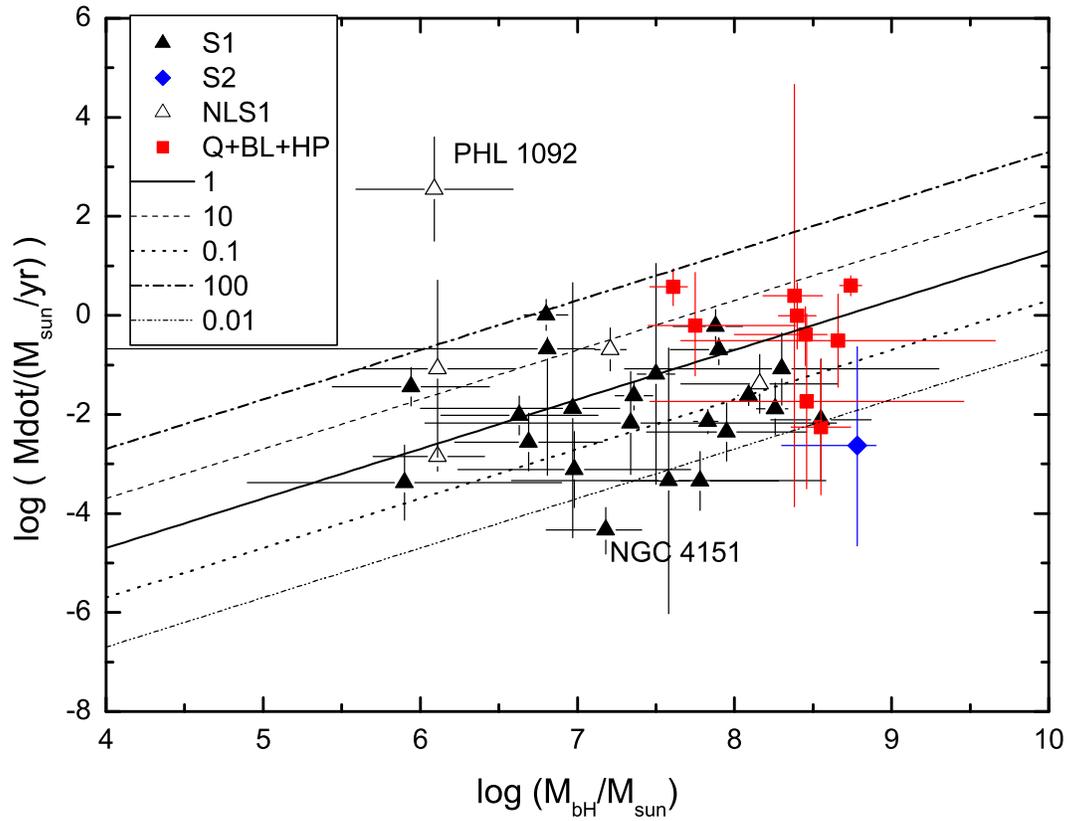}
%\centerline{\epsfig{file=mdot-m.eps,width=10cm,angle=0,clip=}}
\caption{$log(\dot M)$ versus $log(M_{BH})$.  $\dot M = \epsilon
\dot M_{EDD}$. $\epsilon$ is 1, 10, 0.1, 100, 0.01 respectively
for the solid line, dash line, dot line, dash dot line and dash
dot dot line.} \label{mdot-m}
\end{figure}

The accretion rate $\dot M$ can be derived from the relation,
$L_{x}\simeq \frac{GM_{BH}\dot M}{R_{x}}$

\begin{equation}
\dot M=2L_{x}(R_{x}/R_{g})/c^{2}. \label{dotm}
\end{equation}

where $L_{x}$ is the soft X-ray luminosity in 0.1-2.4 keV band, G
is gravitational constant, and $R_{x}$ is the size of the X-ray
radiation region. Using $R_{x}/R_{g}$ and $L_{x}$, accretion rates
for 37 AGNs are obtained from Eq. \ref{dotm}, which are listed in
column 6 of Table 1. The $\dot M$ distribution is $<log(\dot
M)>=-1.50\pm 0.23$ with a standard deviation of 1.42. In Figure
\ref{mdot-m} we show the accretion rates ($\dot M$) versus black
hole masses ($M_{BH}$). The accretion rates of most of AGNs are
100$ \sim $0.01 $\dot M_{Edd}$, taking $\dot
M_{Edd}=0.2\frac{M}{10^{8}\Msolar} (\Msolar/yr)$ .

\begin{figure}
\plotone{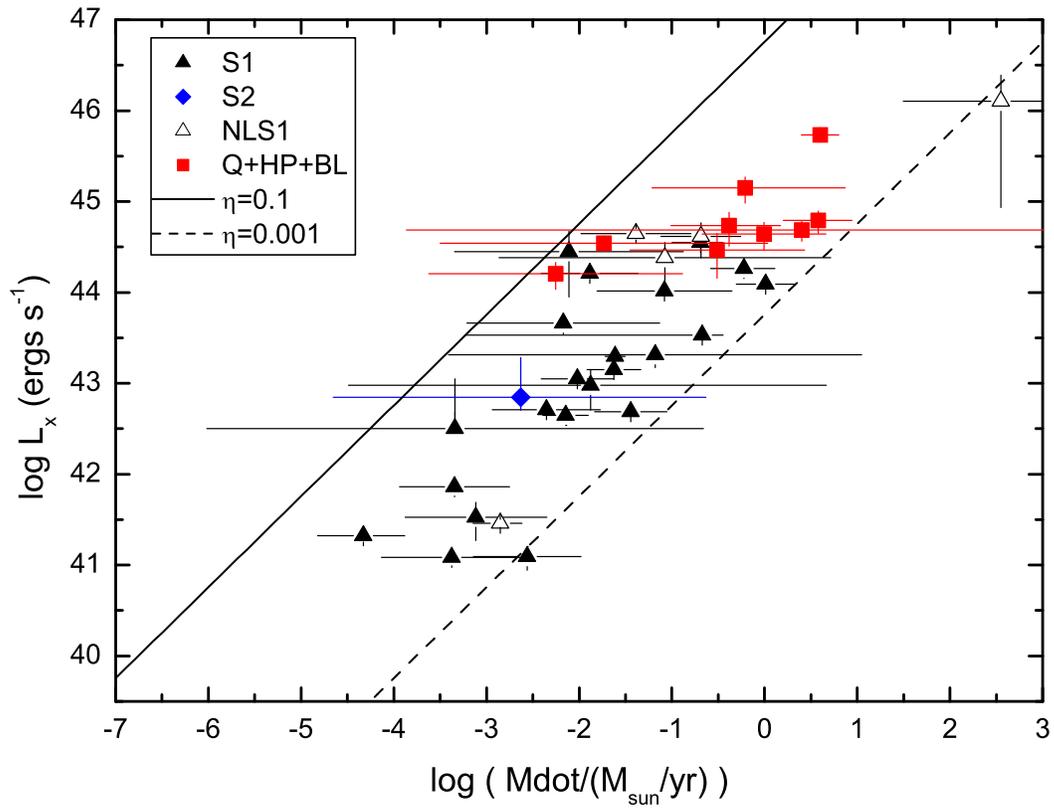}
%\centerline{\epsfig{file=lx-mdot.eps,width=10cm,angle=0,clip=}}
\caption{$log(L_{x})$ versus $log(\dot M)$.  The solid line and
dash line show $L_{x}=\zeta \dot Mc^{2}$, $\zeta=0.1$,
$\zeta=0.001$, respectively.} \label{lx-mdot}
\end{figure}

We plot $L_{x}$ versus $\dot M$ in Figure \ref{lx-mdot}.
$L_{x}=\zeta \dot M c^{2} $, where $\zeta$ is the accretion
efficiency. The solid line in Figure \ref{lx-mdot} is $L_{x}=0.1
\dot M c^{2} $ and the dash line is $L_{x}=0.001 \dot M c^{2} $.
Most AGNs are in the range of 0.1$\sim$0.001 of $\zeta$. Combining
Eq. \ref{dotm}, we can drive the accretion efficiency
$\zeta=1/(2R_{x}/R_{g})$. From Figure \ref{rx-m} we find the
values of $R_{x}/R_{g}$ in most AGNs are in the range of 5$ \sim
$500, which is consistent with the value of $\zeta$ (0.1$ \sim
$0.001).

\subsection{The Compactness Parameter}

\begin{figure}
\plotone{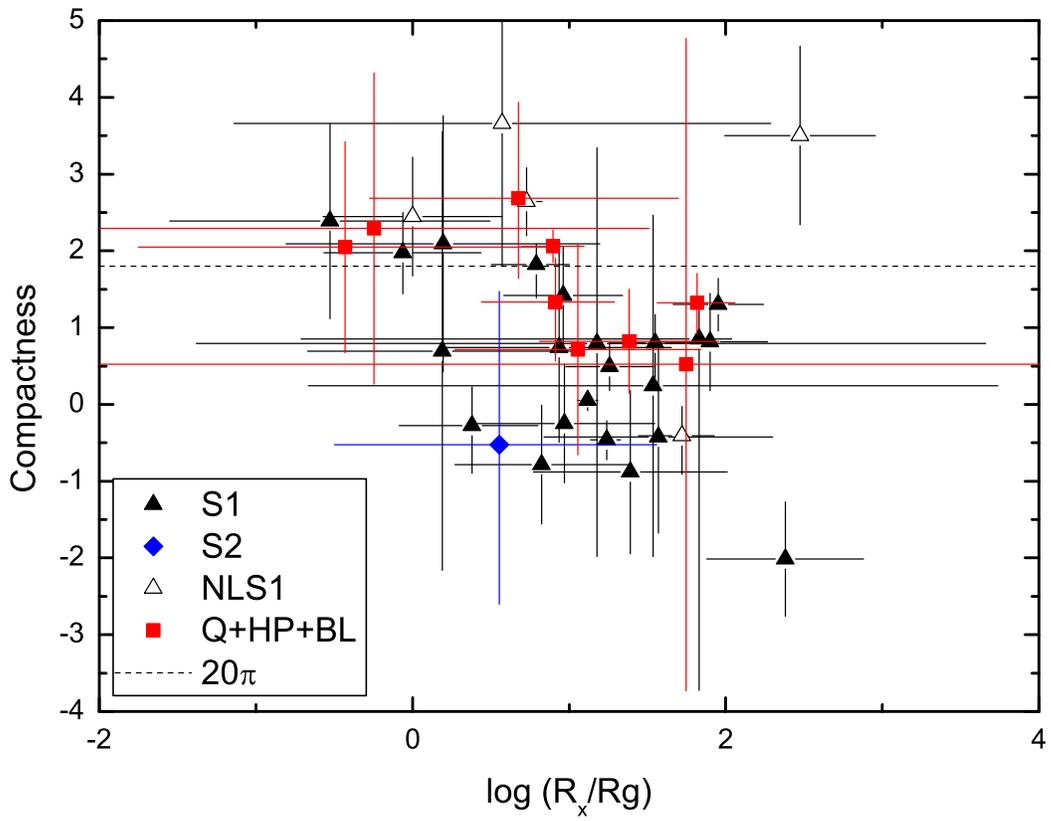}
%\centerline{\epsfig{file=comp-rx.eps,width=10cm,angle=0,clip=}}
\caption{Log of the compactness parameter $l$ versus
$log(R_{x}/R_{g})$. The dash line is $l=20\pi$. } \label{comp-rx}
\end{figure}

If the power law spectra observed in AGNs in the X-ray extend to
much higher energies, then pairs may be expected to be produced if
the optical depth of the $\gamma$-ray photons for pair production
exceeds unity. The condition is usually stated in terms of the
compactness parameter (Done \& Fabian 1989),
\begin{equation}
l=\frac{\sigma_{T}}{m_{e}c^{3}}\frac{L_{x}}{R_{x}}.
\end{equation}
Where $L_{x}$ is the luminosity produced in the region of $R_{x}$,
$m_{e}$ is the mass of electron and $\sigma_{T}$ is Thomson cross
section. The values of the compactness parameters are listed in
column 9 of Table 1. If $l>20\pi$, the source has unity optical
depth in X-ray for $\gamma$-ray absorption at about one Mev and a
significant fraction of the source luminosity can then pass
through electron-positron pairs. The distribution of the
compactness parameter is $<log(l)>=0.90\pm 0.20$ with a standard
deviation of 1.23. We plot the compactness parameter versus
$R_{x}/R_{g}$ in Figure \ref{comp-rx}. A simple least square
linear regression (Press et al. 1992) gives $log(l)=(1.70\pm
0.29)+(-0.85\pm0.25)log(R_{x}/R_{g})$ (excluding PHL 1092)
(R=-0.50, P=0.00168 ). Those AGNs with smaller $R_{x}/R_{g}$
appear to have larger compactness parameter. From Figure
\ref{comp-rx} we can find that several AGNs have $l>20\pi$, while
several AGNs have $l<0.1$, which is consistent with the results of
Done \& Fabian (1989).

\section{THE SIZES OF THE BLR}
Broad emission lines are one of the dominant features of spectra
of AGNs. Broad Line Regions (BLRs) play a particularly important
role in our understanding of AGNs by virtue of their proximity to
the central source. With reverberation mapping techniques, the
sizes of the BLRs can be obtained through the study of correlated
variations of the lines and continuum fluxes (Peterson 1993). The
BLR sizes for 17 Seyfert 1 galaxies (Wandel et al. 1999) and for
17 PG quasars (Kaspi et al. 2000) have been recently obtained. In
Table 2, we list 12 AGNs with available BLRs sizes and the
calculated accretion rates. Here we adopted 3 light days as the
error of the BLRs size of NGC3516, which is not given in Ho
(1998). Errors of BLRs sizes of other 11 AGNs are from Kaspi et
al. (2000).

We assume that the gravitational instability of the standard thin
disk at large radius leads to the formation of BLRs (Bian \& Zhao
2002). The criterion is $Q=\frac{\Omega^{2}}{\pi G\rho}\leq 1$
(Golreich \& Lynden-Bell 1965), where $\Omega$ is the Keplerian
angular velocity at R away from the central black hole
$(GM/R^{3})^{1/2}$, $M$ is the black hole mass, and $\rho$ is the
local mass density. We adopted the solutions of $\rho$ for the
standard thin disk from Shakura \& Sunyaev (1973) ,
$\rho=3.1\times10^{-5}\alpha^{-7/10}\dot{M}_{26}^{11/20}
M_{8}^{5/8}R_{14}^{-15/8}f^{11/5} \rm{~~g~cm^{-3}}$, where
$\alpha$ is the parameter of the standard $\alpha$ disk,
$f=(1-(R/R_{s})^{1/2})^{1/4}$, $\dot{M}_{26}= \dot{M}/(10 ^{26}
\rm{g~s^{-1}})$, $M_{8}=M/(10^{8}\Msolar)$, and
$R_{14}=R/(10^{14}cm)$. $f=1$ and $\alpha=1$ are adopted in AGNs.
We can obtain the sizes of the BLRs if we know the central black
hole masses and accretion rates,
\begin{equation}
R_{14}=880\alpha^{28/45}Q^{-8/9}\dot{M}_{26}^{-22/45}M_{8}^{1/3}.
\end{equation}

The BLRs sizes of 12 AGNs are obtained from Eq. (12), which are
listed in Table 2. The uncertainties of our calculated BLRs sizes
are estimated by considering $1/4 \leq Q \leq 4$ ( Golreich \&
Lynden-Bell 1965 ), errors of the black hole masses and the
accretion rates.

\begin{figure}
\plotone{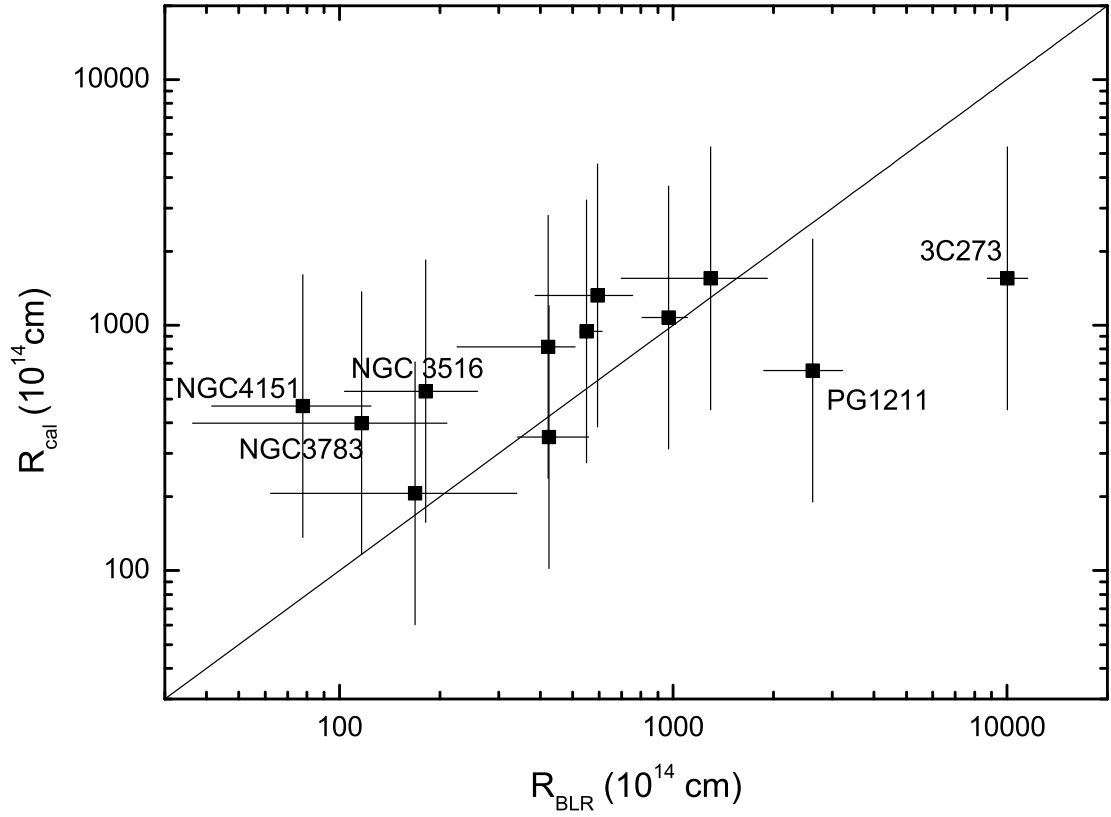}
%\centerline{\epsfig{file=rcal-r.eps,width=10cm,angle=0,clip=}}
\caption{$Log(R_{cal})$ versus $log(R_{BLR})$. The solid line
shows 1:1.}

\label{rcal-r}

\end{figure}

Figure \ref{rcal-r} shows that the calculated BLRs sizes derived
from Eq. 12 are consistent with the sizes from the reverberation
mapping method. We have a $\chi^{2}$ (Press et al. 1992) test of
$logR_{BLR}$ to show whether our calculated BLRs sizes are
consistent with that from reverberation mapping method .
$\chi^{2}$ and probability are 0.933 and 99.996\% for 12 AGNs.

\section{SUMMARY}

We use the exponential time-scale to characterize the soft X-ray
variabilities of AGNs with obvious exponentially increasing or
decreasing patterns in the light curves and obtain time-scales for
65 AGNs. The exponential time-scale is suitable for analyzing
X-ray variability, especially for the data with larger time gaps
such as ROSAT data. The exponential time-scales can provide good
upper limits for the central black hole masses and the sizes of
the X-ray radiation region. From the relationship of $L_{x}$ and
$\Delta t_{e}$ we can reject bremsstrahlung as the primary
radiation source for most AGNs here. Many AGNs have very efficient
engines. There is no explicitly grouped distribution in $\dot M-M$
diagram for different kinds of AGNs. This means that, under the
same soft X-ray luminosity, the central mass or radiation region,
compared with Schwarzschild radius, varies from one AGN to
another.

Using the recent mass estimates of central black holes and the
exponential time-scales of AGNs from ROSAT/PSPC pointing data, the
sizes of X-ray radiation region, the accretion rate and the
compactness parameter of 37 AGNs are obtained. We use the
calculated accretion rates to calculate the radius of
gravitational instability for 12 AGNs among 37 AGNs, which is
consistent with the BLRs sizes from the reverberation mapping
method. This provides further evidence for the gravitational
instability leading to the formation of BLRs.

\acknowledgements

We thank Keliang Huang for useful discussions, and the anonymous
referee for the valuable comments. We thank Helmut Abt for reading
our manuscript. We thank the Chinese Natural Science Foundation
for financial support under contract 10273007.

\clearpage

\clearpage

\begin{table*}

\begin{center}
\begin{tiny}
\begin{tabular}{lllllllll}
\hline \hline

Name &Type& $\Delta T_{e}$& $L_{x}$&$log(M/ \Msolar)$&$log \dot M$&$log(R_{x}/R_{g})$&log(l) \\
     &    &  ($10^{4}sec$)&($10^{43}ergs s^{-1}$)& &($\Msolar~yr^{-1}$)&&\\
(1) & (2)&(3)&(4)&(5)&(6)&(7)&(8)\\
\hline
MARK  335$^{a}$ &  S1   &   $7.45   \pm3.05  $&$27.7  \pm 1.5      $& $6.8         ^{+0.14}_{-0.14}$&$0.011^{+0.31}_{-0.31}$&$1.95       ^{+0.29}_{-0.29}$&$1.30 ^{+0.34 }_{-0.34}$  \\
I Zw 1$^{d}$      &NLS1 &   $0.279  \pm0.302 $&$93    \pm 17.3     $& $7.21        ^{+0.1 }_{-0.1 }$&$-0.69^{+0.43}_{-0.43}$&$0.73       ^{+0.10}_{-0.10}$&$2.64 ^{+0.44 }_{-0.44}$  \\
TON S180$^{c}$  &  NLS1 &   $0.201  \pm0.140^{\star} $&$99.4  \pm 6.5      $& $8.16        ^{+0.5 }_{-0.5 }$&$-1.39^{+0.59}_{-0.59}$&$-1.3E-5    ^{+0.57}_{-0.57}$&$2.44 ^{+0.77 }_{-0.77}$  \\
F   9$^{a}$     &  S1   &   $1.69   \pm0.50  $&$79.9  \pm 3.9$& $7.9         ^{+0.11}_{-0.31}$&$-0.69^{+0.24}_{-0.30}$&$0.79   ^{+0.21}_{-0.29}$&$1.82 ^{+0.261}_{-0.44}$  \\
PHL 1092$^{f}$  &  NLS1 &   $8.84   \pm4.60  $&$2870  \pm 1190     $& $6.09        ^{+0.5 }_{-0.5 }$&$2.54 ^{+1.05}_{-1.05}$&$2.48       ^{+0.48}_{-0.48}$&$3.50 ^{+1.16 }_{-1.16}$  \\
MS 01585+0019   &  BL   &   $0.0529 \pm0.0748^{\star}$&$274   \pm 57       $& $                            $&$                     $&$                           $&$                      $  \\
MARK  586$^{g}$ &  Q    &   $0.806  \pm1.124 $&$317   \pm 45       $& $7.75        ^{+0.64}_{-0.30}$&$-0.21^{+1.08}_{-1.01}$&$0.68       ^{+1.02}_{-0.95}$&$2.69 ^{+1.25 }_{-1.05}$  \\
ESO 198-G24     &  S1   &   $1.58   \pm0.74^{\star}  $&$26.3  \pm 1.4      $& $                            $&$                     $&$                           $&$                      $  \\
MARK  372       &  S1   &   $8.11   \pm12.37 $&$4.11  \pm 0.60     $& $                            $&$                     $&$                           $&$                      $  \\
NGC 1275$^{b}$  &  S1.5 &   $0.08   \pm0.122^{\star} $&$63    \pm 19.1     $& $ 8.55       ^{+0.32}_{-0.32}$&$-2.11^{+1.23}_{-1.23}$&$-0.53      ^{+1.02}_{-1.02}$&$2.39 ^{+1.27 }_{-1.27}$  \\
NGC 1566$^{b}$  &  S1   &   $25.2   \pm14.8  $&$0.0278\pm 0.0036   $& $6.69        ^{+0.47}_{-0.47}$&$-2.56^{+0.58}_{-0.58}$&$2.38   ^{+0.50}_{-0.50}$&$-2.01^{+0.745}_{-0.75}$  \\
AKN  120$^{a}$  &  S1   &   $0.204  \pm0.152 $&$36.4  \pm 2.2      $& $8.26        ^{+0.08}_{-0.12}$&$-1.89^{+0.52}_{-0.52}$&$-0.06      ^{+0.50}_{-0.50}$&$1.97 ^{+0.52 }_{-0.53}$  \\
PKS 0548-322    &  HP   &   $0.0745 \pm0.1561$&$150   \pm 5        $& $                            $&$                     $&$                           $&$                      $  \\
S5 0716+71      &  HP   &   $0.147  \pm0.052^{\star} $&$                   $& $                            $&$                     $&$                           $&$                      $  \\
MS 07379+7441   &  BL   &   $2.69   \pm2.60^{\star}  $&$861    \pm 95       $& $                            $&$                     $&$                           $&$                      $  \\
MARK   10$^{b}$ &  S1   &   $0.0596 \pm0.0438$&$10.4  \pm 1.2      $& $7.34        ^{+1.31}_{-1.31}$&$-2.17^{+1.04}_{-1.04}$&$0.19       ^{+1.00}_{-1.00}$&$2.09 ^{+1.67 }_{-1.67}$  \\
VII Zw 244      &  S1   &   $1.12   \pm0.52^{\star}  $&$62.7   \pm 18.9     $& $                            $&$                     $&$                           $&$                      $  \\
ESO 434-G40$^{h}$&  S1 &  $0.102   \pm0.083 $&$0.715 \pm 0.806    $& $7.58        ^{+1   }_{-1   }$&$-3.34^{+2.68}_{-2.68}$&$0.19   ^{+0.86}_{-0.86}$&$0.69 ^{+2.868}_{-2.86}$  \\
NGC 3031(M81)$^{e}$& S1 &   $6.82   \pm0.86  $&$                   $& $7.83        ^{+0.04}_{-0.09}$&$-2.14^{+0.24}_{-0.25}$&$1.24       ^{+0.09}_{-0.10}$&$-0.46^{+0.24 }_{-0.26}$  \\
IRAS 09595-075  &  S1   &   $41.0   \pm14.4  $&$13.1  \pm 1.4      $& $                            $&$                     $&$                           $&$                      $  \\
HE 1029-1401    &  Q    &   $4.73   \pm3.39  $&$236   \pm 12       $& $                            $&$                     $&$                           $&$                      $  \\
MS 10590+7302   &  S1   &   $0.904  \pm0.889^{\star} $&$18.7  \pm 2.7      $& $                            $&$                     $&$                           $&$                      $  \\
NGC 3516$^{a}$  &  S1   &   $2.46   \pm0.99  $&$3.17  \pm 0.1      $& $7.36        ^{+0.12}_{-0.12}$&$-1.62^{+0.29}_{-0.29}$&$1.26       ^{+0.28}_{-0.27}$&$0.49 ^{+0.31 }_{-0.31}$  \\
MARK  180$^{b}$ &  BL   &   $0.112  \pm0.221 $&$36.2  \pm 5.3      $& $8.55        ^{+0.19}_{-0.19}$&$-2.25^{+1.37}_{-1.37}$&$-0.43      ^{+1.32}_{-1.32}$&$2.05 ^{+1.38 }_{-1.38}$  \\
NGC 3783$^{a}$  &  S1   &   $0.756  \pm2.806 $&$2.15  \pm 0.45     $& $6.97        ^{+0.30}_{-0.97}$&$-1.87^{+2.54}_{-2.61}$&$1.18   ^{+2.48}_{-2.56}$&$0.80 ^{+2.544}_{-2.78}$  \\
MARK 1310       &  S1   &   $0.529  \pm0.366^{\star} $&$0.451 \pm 0.06     $& $                            $&$                     $&$                           $&$                      $  \\
NGC 4051$^{a}$  &  NLS1 &   $0.684  \pm0.05^{\star}  $&$0.0651\pm 0.003    $& $ 6.11       ^{+0.30}_{-0.41}$&$-2.85^{+0.23}_{-0.30}$&$1.72       ^{+0.21}_{-0.27}$&$-0.4 ^{+0.38 }_{-0.51}$  \\
GQ COM          &  Q    &   $0.0713 \pm0.0575$&$72.4  \pm 7.1      $& $                            $&$                     $&$                           $&$                      $  \\
NGC 4151$^{a}$  &  S1   &   $0.0784 \pm0.0461$&$0.0472\pm 0.0031   $& $7.18         ^{+0.23}_{-0.38}$&$-4.33^{+0.44}_{-0.49}$&$0.38       ^{+0.42}_{-0.46}$&$-0.2 ^{+0.50 }_{-0.62}$  \\
PG 1211+143     &  Q    &   $30.2   \pm10.6^{\star}  $&$139   \pm 17       $& $7.61        ^{+0.09}_{-0.15}$&$0.577^{+0.37}_{-0.38}$&$1.82       ^{+0.24}_{-0.24}$&$1.32 ^{+0.38 }_{-0.40}$  \\
B2 1215+30      &  HP   &   $10.4   \pm4.6   $&$                   $& $                            $&$                     $&$                           $&$                      $  \\
MARK  766$^{c}$ &  S1   &   $0.165  \pm0.045^{\star} $&$2.52  \pm 0.09     $& $ 6.63       ^{+0.5 }_{-0.5 }$&$-2.02^{+0.39}_{-0.39}$&$0.96       ^{+0.38}_{-0.38}$&$1.42 ^{+0.63 }_{-0.63}$  \\
PG 1307+085$^{a}$&  Q   &   $9.19   \pm5.03  $&$122   \pm 22       $& $8.45        ^{+0.14}_{-0.45}$&$-0.38^{+0.55}_{-0.62}$&$0.91       ^{+0.38}_{-0.47}$&$1.33 ^{+0.57 }_{-0.77}$  \\
PG 1218+304     &  HP   &   $6.61   \pm2.21  $&$                   $& $                            $&$                     $&$                           $&$                      $  \\
B2 1223+25$^{h}$ &  Q   &   $24.3   \pm15.1  $&$65.9  \pm 15       $& $ 8.66                       $&$-0.51^{+0.94}_{-0.94}$&$1.05       ^{+0.78}_{-0.78}$&$0.71 ^{+1.37 }_{-1.37}$  \\
3C 273.0$^{a}$  &  Q    &   $16.9   \pm4.9^{\star}   $&$1220  \pm 20       $& $8.74        ^{+0.07}_{-0.07}$&$ 0.60^{+0.20}_{-0.20}$&$0.90       ^{+0.20}_{-0.20}$&$2.06 ^{+0.21 }_{-0.21}$  \\
TON 1542$^{a}$  &  S1   &   $22.2   \pm7.6   $&$41.2  \pm 3.9      $& $ 7.88       ^{+0.17}_{-0.27}$&$-0.22^{+0.33}_{-0.36}$&$1.55       ^{+0.25}_{-0.29}$&$0.80 ^{+0.37 }_{-0.45}$  \\
NGC 4579$^{b}$    &S1.9 &   $1.45   \pm0.96^{\star}  $&$0.163 \pm 0.015    $& $7.78        ^{+0.5 }_{-0.5 }$&$-3.34^{+0.59}_{-0.59}$&$0.83       ^{+0.55}_{-0.55}$&$-0.78^{+0.77 }_{-0.77}$  \\
WAS 61          &  NLS1 &   $10.8   \pm0.44^{\star}  $&$12.4  \pm 1.8      $& $                            $&$                     $&$                           $&$                      $  \\
PG 1244+026$^{c}$ &NLS1 &   $0.0130 \pm0.0328$&$54.1  \pm 11.5     $& $6.11        ^{+0.5 }_{-0.5 }$&$-1.08^{+1.79}_{-1.79}$&$0.57       ^{+1.71}_{-1.71}$&$3.66 ^{+1.84 }_{-1.85}$  \\
MS 12480-0600A  &  S1   &   $0.222  \pm0.714 $&$110   \pm 19       $& $                            $&$                     $&$                           $&$                      $  \\
K07.01          &  S1   &   $1.33   \pm1.19^{\star}  $&$31.8  \pm 9.1      $& $                            $&$                     $&$                           $&$                      $  \\
NGC 5033$^{h}$  &  S1.9 &   $0.251  \pm0.122^{\star} $&$0.0274\pm 0.0024   $& $5.9         ^{+1   }_{-1   }$&$-3.38^{+0.76}_{-0.76}$&$1.57       ^{+0.73}_{-0.73}$&$-0.4 ^{+1.25 }_{-1.25}$  \\
K08.02          &  S1   &   $0.0557 \pm0.0354^{\star}$&$13.1  \pm 2.2      $& $                            $&$                     $&$                           $&$                      $  \\
MS 13326-2935   &  BL   &   $4.39   \pm2.45  $&$87.2  \pm 15.3     $& $                            $&$                     $&$                           $&$                      $  \\
MCG 06.30.015$^{f}$ & S1&   $0.855  \pm0.197 $&$1.10  \pm 0.06     $& $5.94        ^{+0.5 }_{-0.5 }$&$-1.44^{+0.39}_{-0.39}$&$1.90   ^{+0.37}_{-0.37}$&$0.81 ^{+0.632}_{-0.63}$  \\
IRAS 13349+243  &  Q    &   $559    \pm118   $&$82.8  \pm 15.2     $& $                            $&$                     $&$                           $&$                      $  \\
2E 1346+2646    &  NLS1 &   $19     \pm5.8^{\star}   $&$4.53  \pm 0.65     $& $                            $&$                     $&$                           $&$                      $  \\
NGC 5548  $^{a}$ &  S1  &   $8.17   \pm0.26 ^{\star} $&$4.47  \pm 0.08     $& $ 8.09       ^{+0.07}_{-0.07}$&$-1.61^{+0.07}_{-0.07}$&$1.12       ^{+0.05}_{-0.05}$&$0.05 ^{+0.10 }_{-0.10}$  \\
PG 1416-129$^{d}$ &S1.2 &   $41.5   \pm35.5  $&$99    \pm 14.7     $& $8.40        ^{+0.12}_{-0.12}$&$-0.01^{+0.66}_{-0.67}$&$1.38   ^{+0.57}_{-0.57}$&$0.82 ^{+0.67 }_{-0.67}$  \\
MARK  474       &  S1   &   $1.54   \pm0.66^{\star}  $&$13.6  \pm 1        $& $                            $&$                     $&$                           $&$                      $  \\
MARK  841       &  S1   &   $7.05   \pm2.81  $&$23.3  \pm 1        $& $8.3                         $&$-1.08^{+0.73}_{-0.73}$&$0.94       ^{+0.72}_{-0.72}$&$0.74 ^{+1.23 }_{-1.23}$  \\
MARK  290$^{d}$ &  S1   &   $8.86   \pm29.24^{\star} $&$4.64  \pm 0.57     $& $7.50        ^{+0.12}_{-0.12}$&$-1.18^{+2.23}_{-2.23}$&$1.54       ^{+2.20}_{-2.20}$&$0.24 ^{+2.22 }_{-2.22}$  \\
MARK  876$^{a}$ &  Q    &   $139    \pm88.3^{\star}  $&$109   \pm 12       $& $8.38        ^{+0.18}_{-0.20}$&$ 0.40^{+4.27}_{-4.27}$&$1.75       ^{+4.24}_{-4.23}$&$0.52 ^{+4.25 }_{-4.25}$  \\
NGC 6251$^{e}$  &  S2   &   $5.69   \pm8.55^{\star}  $&$1.58   \pm 1.21     $& $8.78        ^{+0.12}_{-0.48}$&$-2.63^{+2.00}_{-2.02}$&$0.55       ^{+1.00}_{-1.05}$&$-0.53^{+2.00 }_{-2.07}$  \\
3C 345.0        &  HP   &   $2260   \pm1180^{\star}  $&$644   \pm 51       $& $                            $&$                     $&$                           $&$                      $  \\
EXO 1652.4+393  &  NLS1 &   $0.075  \pm0.0567^{\star}$&$9.18  \pm 3.28     $& $                            $&$                     $&$                           $&$                      $  \\
NGC 6814$^{b}$  &  S1.5 &   $1.62   \pm0.91  $&$0.0758\pm 0.015    $& $6.98        ^{+0.74}_{-0.74}$&$-3.11^{+0.77}_{-0.77}$&$1.39   ^{+0.62}_{-0.62}$&$-0.88^{+1.06 }_{-1.06}$  \\
F  339          &  S1   &   $6.12   \pm2.54  $&$81.2  \pm 18.4     $& $                            $&$                     $&$                           $&$                      $  \\
NGC 7214        &  S1   &   $1.61   \pm0.76^{\star}  $&$7.35  \pm 0.38     $& $                            $&$                     $&$                           $&$                      $  \\
NGC 7213$^{b}$  &  S1   &   $3.53   \pm2.47^{\star}  $&$1.15  \pm 0.03     $& $7.95        ^{+0.51}_{-0.51}$&$-2.35^{+0.58}_{-0.58}$&$0.97   ^{+0.58}_{-0.58}$&$-0.25^{+0.772}_{-0.77}$  \\
MS 22549-3712   &  S1   &   $0.341  \pm0.293 $&$16.5  \pm 3.6      $& $                            $&$                     $&$                           $&$                      $  \\
NGC 7469$^{a}$  &  S1   &   $4.99   \pm0.40^{\star}  $&$7.63  \pm 0.28     $& $6.81        ^{+0.30}_{-3.81}$&$-0.67^{+0.22}_{-2.55}$&$1.83       ^{+0.21}_{-2.54}$&$0.85 ^{+0.37 }_{-4.58}$  \\
MARK  926$^{h}$  & S1.5 &   $0.172  \pm0.419^{\star} $&$78.6  \pm 4.5      $& $8.46      ^{+1   }_{-1   }$&$-1.73^{+1.77}_{-1.77}$&$-0.2 ^{+1.76}_{-1.76}$&$2.29 ^{+2.02 }_{-2.02}$  \\
MS 23409-1511   &  NLS1 &   $0.126  \pm0.192 $&$249   \pm 32       $& $                            $&$                     $&$                           $&$                      $  \\

\hline
\end{tabular}
\caption{The properities of 65 AGNs in Right Ascension order.
Col.1: name, Col.2: type, Col.3: the exponential time-scale,
Col.4: 0.1-2.4 keV Luminosity , Col.5:log of the BH mass in
$\Msolar$, Col.6: log of calculating accretion rates in
$\Msolar/yr$ , Col.7 log of the ratio of the size of X-ray
emission region and the Schwarzschild radius , Col.8: log of the
compactness parameter. $^{\star}$: the exponential time-scale from
the rising variation. Estimation of central black hole masses: a:
Nelson 2000; b: Nelson \& Whittle 1995; Falomo, et al. 2002; c:
Wang \& Lu 2001; d: Vestergaard 2002; e: Tremaine et al. 2002; f:
Czerny et al. 2001; g: Mathur et al. 2001; h: Padovani et al.
1990.}

\end{tiny}
\end{center}
\end{table*}

\begin{table*}
\begin{center}

\begin{tabular}{lllll}
\hline
\hline
Name &$R_{BLR}$&$log(M/ \Msolar)$&$log \dot M$&$R_{cal}$ \\
     & $(10^{14}cm)$& &$(\Msolar~yr^{-1})$ & $(10^{14}cm)$ \\
\hline
F9       &$  422.5^{+85.5}_{-197}$&$  7.90 ^{+0.11 }_{-0.31}$&       $ -0.69  ^{+ 0.24  }_{- 0.31 }  $  & $   815.0 ^{+  1979.6}_{-577.3}$ \\
NGC3783  &$  116.7^{+93.3}_{-80.4}$&$  6.97^{+0.30  }_{-0.97}$&      $ -1.87 ^{+ 2.54 }_{- 2.61 }  $  & $      399.2^{+  969.6}_{-282.8}  $ \\
AKN120   &$  969.4^{+132.2}_{-163.3}$&$  8.26^{+0.08 }_{-0.12}$&    $ -1.89 ^{+ 0.52 }_{- 0.52}  $  & $      1074.4^{+  2609.6}_{-761.0}   $\\
3C273    &$  10031^{+1503.4}_{-1296 }$&$  8.74^{+0.07 }_{-0.07}$&   $ ~0.60  ^{+ 0.20 }_{- 0.20}  $  & $      1552.9^{+  3772.0}_{-1100.0}  $ \\
PG1211   &$ 2617.9^{+596.2}_{-751.7 }$&$  7.61^{+0.09 }_{-0.15}$&   $ ~0.58  ^{+ 0.37 }_{- 0.38}  $  & $       652.4^{+  1584.5}_{-462.1}    $ \\
MARK335  &$  425.1^{+132.2}_{-82.9}$&$  6.80 ^{+0.14 }_{-0.14}$&     $ ~0.01 ^{+  0.31}_{- 0.31} $  & $         350.3^{+  851.0}_{-248.2}      $ \\
Ton1542  &$  1296^{+622.1}_{-596.2}$&$  8.74^{+0.17 }_{-0.27}$&     $  -0.22^{+  0.33}_{- 0.36} $  & $       1552.9^{+  3772.0}_{-1100.0}   $\\
NGC5548  &$  549.5^{+62.2}_{-18.1}$&$  8.09^{+0.07 }_{-0.07}$&      $ -1.61 ^{+  0.065}_{- 0.065} $  & $      942.9^{+  2290.4}_{-667.9}   $\\
NGC4151  &$   77.8^{+46.7}_{-36.3}$&$  7.18^{+0.23 }_{-0.38}$&      $ -4.33 ^{+  0.445 }_{- 0.49 } $  & $     469.0^{+  1139.1}_{-332}  $ \\
NGC4051  &$  168.5^{+171.1}_{-106.3}$&$  6.11^{+0.30 }_{-0.41}$&    $ -2.85 ^{+  0.23}_{- 0.30} $  & $        206.3^{+  501.1}_{-146}      $\\
3C390.3  &$  593.6^{+163.3}_{-207.4}$&$  7.36^{+0.10  }_{-0.10 }$&    $ -1.62 ^{+ 0.29 }_{- 0.29} $  & $       1321.7^{+  3210.5}_{-936.3}   $\\
NGC3516  &$  181.4^{+77.8}_{-77.8}$&$  8.53^{+0.12 }_{-0.21}$&      $  ~0.81 ^{+ 0.21 }_{- 0.66}  $  & $       538.5^{+  1307.9}_{-381.4}    $\\
\hline

\end{tabular}
\caption{The BLRs size  of 12 AGNs. Col.1: name, Col.2: BLRs sizes
from reverberation mapping method (Kaspi  et al. 2000) , Col.3:
central black hole masses (Kaspi et al. 2000) , Col.4: calculating
accretion rate, Col. 5: calculating BLRs size. }
\end{center}
\end{table*}

\end{document}